\documentclass[journal]{IEEEtran}

\ifCLASSINFOpdf
\else
   \usepackage[dvips]{graphicx}
\fi
\usepackage{url}

\usepackage{graphicx}

\usepackage{multirow}
\usepackage{adjustbox}
\usepackage{amsmath}
\usepackage{color}
\usepackage{amssymb}
\usepackage{hyperref}
\usepackage{booktabs}
\usepackage{makecell}
\usepackage{array}
\usepackage{caption}
\usepackage{subcaption}  
\usepackage{tabularx}
\begin{document}

\title{Preference Optimization for Non-Verbal Vocalization Synthesis}

\author{Haoyang Li, Chenglin Xu, Junchuan Zhao, Yuang Cao, Liumeng Xue, Yiwen Guo, Eng Siong Chng

\thanks{Haoyang Li, Eng Siong Chng are with Nanyang Technological University, Singapore (e-mail: LI0078NG@e.ntu.edu.sg; ASESChng@ntu.edu.sg;). Chenglin Xu is with LIGHTSPEED, Singapore (e-mail: cxu011@e.ntu.edu.sg;). Junchuan Zhao is with National University of Singapore, Singapore (e-mail: junchuan@u.nus.edu;). Yuang Cao, Liumeng Xue are with Nanjing University, China (e-mail: cya696969@gmail.com; lmxue@nju.edu.cn;). Yiwen Guo is an independent researcher (e-mail: guoyiwen89@gmail.com;). Correspondence to Chenglin Xu. This work was done during Haoyang Li's research internship at LIGHTSPEED.}}

\markboth{Journal of \LaTeX\ Class Files, Vol. 14, No. 8, August 2015}
{Shell \MakeLowercase{\textit{et al.}}: Bare Demo of IEEEtran.cls for IEEE Journals}
\maketitle

\begin{abstract}
Non-verbal vocalizations (NVs), such as laughter, coughs, and sighs, are essential for expressive TTS, but the effectiveness of preference optimization for NV generation remains poorly understood. We systematically study preference optimization for NV-capable TTS, focusing on preference signals, preference-pair construction, and DPO-based optimization objectives. We formulate an NV-aware character error rate (NV-CER) by treating NV tags as distinct output symbols and computing a weighted pinyin-based CER over both verbal and non-verbal content, enabling controllable optimization of NV realization without modifying the underlying optimization algorithm. Experiments on Emilia-NV and the augmented NV-Bench covering 18 NV types reveal how different design choices affect NV realization and lexical fidelity, and establish an effective setup using standard DPO. Objective, LLM-based, and human evaluations provide converging evidence for our findings, offering practical insights into NV-aware post-training for expressive TTS.
\end{abstract}


\IEEEpeerreviewmaketitle

\section{Introduction}
\label{sec:intro}
Recent advances in neural text-to-speech (TTS) have enabled increasingly natural and expressive speech synthesis \cite{shen2024naturalspeech, chen2025f5, hu2026qwen3}. Beyond verbal content, natural communication also relies on non-verbal vocalizations (NVs), including laughter, crying, and hesitation sounds, which convey emotion, discourse structure, and speaker intent \cite{eyben2015geneva}. However, faithful NV generation remains challenging and relatively underexplored in TTS. Recent work has advanced NV-aware TTS through dataset construction \cite{borisov2025nonverbaltts, ye2025scalable, wu2025smiip, liao2026emilia, bai2026synparaspeech} and benchmark development for standardized evaluation \cite{ni2026nv, xue2026nvbench}. These resources enable systematic investigation of NV realization in TTS, an area that has received limited attention to date.

Preference optimization \cite{rafailov2023direct, shao2024deepseekmath} has emerged as an effective post-training paradigm for TTS, improving aspects of naturalness and intelligibility \cite{gao2025emo, tian2025preference, yeo2026improving, yao2025fpo}. Recent TTS systems, including Fish Audio S2 \cite{liao2026fish}, CosyVoice2 \cite{du2024cosyvoice}, and CosyVoice3 \cite{du2025cosyvoice}, have incorporated NV-aware models into preference optimization pipelines. However, these studies primarily focus on overall synthesis quality, providing limited insight into the specific impact of post-training on NV generation. Fish Audio S2 reports final system performance without isolating the contribution of preference optimization, while CosyVoice2 and CosyVoice3 lack NV evaluation. \cite{bai2026synparaspeech} applied DPO to NV-TTS using preference pairs constructed from manually verified speech with NVs and corresponding speech without NVs, encouraging NV occurrence rather than faithful NV realization, and reported degraded CER performance. Consequently, both the effectiveness of preference optimization for NV generation and the design choices that govern its performance remain unclear, motivating a systematic study of preference signals, preference-pair construction, and optimization objectives for NV-aware TTS.

In this work, we systematically investigate NV-aware preference optimization for LLM-based TTS. We formulate an NV-aware character error rate (NV-CER) by treating non-verbal vocalization tags as distinct output symbols and computing a weighted pinyin-based CER over both verbal and non-verbal content. Building on this metric, we construct preference signals using an NV-capable automatic speech recognition (ASR) model and introduce a simple weighting mechanism to adjust the relative importance of NV tags, enabling explicit control over the trade-off between NV realization and lexical fidelity without modifying the underlying optimization algorithm. We then systematically study preference signals, preference-pair construction strategies, and loss formulations within DPO. Our experiments establish how these design choices affect NV realization, leading to an effective setup using standard DPO. Objective, LLM-based, and human evaluations provide converging evidence, establishing practical principles for effective preference optimization and post-training of NV-aware TTS.

\section{Methodology}

\subsection{NV-Aware Preference Signal}
\label{subsec:ps}

\subsubsection{NV-Aware Speech Recognition}

Conventional automatic speech recognition (ASR) systems transcribe only spoken text and disregard non-verbal vocalizations (NVs), making them unsuitable for evaluating expressive speech synthesis. A straightforward alternative is to compare the recognized text against the reference transcription. However, this introduces a fundamental limitation for NV evaluation. Many non-verbal vocalizations, such as \textit{laughter}, \textit{crying}, and \textit{coughing}, do not have deterministic textual representations, making it impossible to establish a one-to-one mapping between speech and text. Although some vocalizations (e.g., hesitation or surprise sounds) may be approximated by lexical words or pinyin, these mappings are inherently ambiguous and context-dependent.

To overcome this limitation, we leverage an NV-aware ASR (NV-ASR) model to recognize predefined NV events alongside spoken text. Given a synthesized speech $x$, the NV-ASR model predicts $\hat{\mathbf{y}}=\{\hat{y}_1,\hat{y}_2,\ldots,\hat{y}_N\}$, where each token represents either a lexical token or an NV tag. Unlike conventional ASR, this transcription preserves explicit NV information for constructing the proposed preference signal.

\subsubsection{NV-Aware Character Error Rate}

Given the reference and predicted transcriptions, we define an \textbf{NV-aware Character Error Rate (NV-CER)} to measure both lexical and NV correctness. For Mandarin, lexical tokens are represented using pinyin to reduce sensitivity to homophones and character variations, while NV tags are retained as independent symbols. Edit distance is computed over the combined token sequence,

\begin{equation}
\mathrm{NV\text{-}CER} = D_c(\mathbf{r},\hat{\mathbf{y}}) / {\scriptstyle\sum_{t\in\mathbf{r}}} c(t)
\label{eq:nv_cer}
\end{equation}

where $\mathbf{r}$ and $\hat{\mathbf{y}}$ denote the reference and predicted token sequences, respectively, and $D_c(\cdot,\cdot)$ denotes the weighted edit distance with token-dependent insertion and deletion costs given by $c(t)$ and substitution cost given by the maximum cost of the two tokens. The proposed framework is readily extendable to other languages by replacing the lexical representation with an appropriate language-specific alternative while preserving NV tags as independent symbols.

\subsubsection{Controllable NV Preference Strength}
\label{sec:nv_weight}

Different applications may require different trade-offs between NV realization and lexical accuracy. To enable this flexibility, we assign a higher edit cost to NV tags, with $w_{\mathrm{NV}}$ denoting the NV weight:

\begin{equation}
c(t)=
\begin{cases}
w_{\mathrm{NV}}, & \text{if } t \text{ is an NV tag},\\
1, & \text{otherwise},
\end{cases}
\label{eq:nv_cer_w}
\end{equation}

Increasing $w_{\mathrm{NV}}$ emphasizes NV errors in NV-CER and thus their influence on preference construction, while $w_{\mathrm{NV}}=1$ assigns equal costs to lexical and NV errors. This weighting provides explicit control over their relative importance without modifying the preference optimization algorithm.

\subsection{NV-aware Preference Alignment}
\label{subsec:pa}

\subsubsection{NV-Aware Preference Pair Construction}
\label{subsubsec: NV-Aware Preference Pair Construction}

We adopt Direct Preference Optimization (DPO) as the preference optimization framework due to its simplicity, effectiveness, and widespread adoption in TTS post-training. Given an input text prompt $p$, the pretrained TTS model independently generates $K$ candidate speech utterances $\mathcal{X}=\{x_1,x_2,\ldots,x_K\}$. Each candidate is transcribed using the NV-ASR model described in Section \ref{subsec:ps}, and its NV-CER score is computed against the reference transcription. The candidate with the lowest NV-CER is selected as the preferred response $x^+$, while the candidate with the highest NV-CER is selected as the rejected response $x^-$, forming the preference pair used for optimization. In addition, we investigate several alternative preference construction strategies, including utilizing ground-truth speech during preference selection and manipulating the synthetic speech. These variants are described in Section \ref{subsubsec: design choices}.

\subsubsection{Preference Optimization Objective}

Given the constructed preference pairs $(x^+,x^-)$, the TTS model is optimized to directly increase the relative likelihood of the preferred response over the rejected response via DPO:

\begin{equation}
\mathcal{L}_{\mathrm{DPO}}
=
-
\log
\sigma
\left(
\beta
\left[
\log
\frac{\pi_\theta(x^+|p)}
{\pi_{\mathrm{ref}}(x^+|p)}
-
\log
\frac{\pi_\theta(x^-|p)}
{\pi_{\mathrm{ref}}(x^-|p)}
\right]
\right),
\label{eq:dpo}
\end{equation}

where $\pi_\theta$ and $\pi_{\mathrm{ref}}$ denote the trainable and reference TTS models, respectively, and $\beta$ controls the preference strength.

Besides the standard DPO objective, we investigate several design choices for NV-aware preference optimization, including combining DPO with the conventional supervised fine-tuning (SFT) objective and alternative preference ranking in Section \ref{subsubsec: design choices}, with their effectiveness evaluated in Section \ref{sec:result}.

\section{Experiments}

\subsection{Dataset}

We train on Emilia-NV ~\cite{liao2026emilia}, comprising 573.4 hours of expressive Mandarin speech with transcriptions annotated by 18 predefined NV tags, covering laughter, breathing, hesitation, and other vocal events.

We evaluate on the Mandarin subset of NV-Bench~\cite{ni2026nv}, which provides text prompts, speaker-conditioning recordings, and ground-truth speech. It contains a single-label subset with 50 utterances per NV category and a 391-utterance multi-label subset, each containing two or more NV events. As NV-Bench covers only 15 of the 18 training tags, we further construct 50 prompts for each of the three missing categories using \texttt{gemini-3.1-pro-preview}, following the NV-Bench annotation style and using its speaker-conditioning utterances. This adds 150 samples to the single-label subset.

\subsection{Evaluation Metrics}
Evaluation is based on recent NV benchmarks \cite{ni2026nv, xue2026nvbench}.

\subsubsection{Objective metrics}

\textbf{Speech intelligibility.}
NV-CER (Eq.~\ref{eq:nv_cer}) evaluates overall transcription accuracy by jointly measuring lexical content and NVs, with the NV weight $w_{\mathrm{NV}}$ set to 1. To separately assess verbal and non-verbal generation, we report CER, computed after removing all NV tags, and PCER, computed after excluding all lexical tokens. We further measure NV retention ratios (the proportion of reference tags correctly preserved) and false positives (FP; count of excess generated tags) for fine-grained per-class NV analysis.

\begin{table}[t]
\centering
\caption{Ablation on loss and preference pair construction. GT: ground-truth speech; $Syn^{+}$/$Syn^{-}$: best-/worst-scoring synthetic candidates; Hybrid: GT or $Syn^{+}$ selected by NV-CER; $Syn^{\mathrm{NoNV}}$: synthetic speech without NVs. Bold/underline: best/second-best.}
\label{tab:ablation_setting}
\resizebox{\linewidth}{!}{
\begin{tabular}{llccc}
\toprule
\textbf{Loss} & \textbf{Pair (Pref, Rej)} & \textbf{NV-CER} & \textbf{PCER} & \textbf{CER} \\
\midrule
DPO & GT, $Syn^-$ & 93.42 & 104.89 & 93.20 \\
DPO+SFT & GT, $Syn^-$ & 9.43  & 39.33  & 8.66  \\
\midrule
DPO & Hybrid, $Syn^-$ & 65.18 & 75.22  & 64.99 \\
DPO+SFT & Hybrid, $Syn^-$ & 7.64  & 33.22  & 7.00  \\
\midrule
DPO & $Syn^+$, $Syn^{NoNV}$ & 22.95 & 52.89 & 22.23 \\
DPO+SFT & $Syn^+$,$Syn^{NoNV}$ & 3.68 & 25.56 & 3.12 \\
\midrule
DPO & $Syn^+$, $Syn^-$ & \textbf{2.59} & \textbf{21.33} & \textbf{2.09} \\
DPO+SFT & $Syn^+$, $Syn^-$ & \underline{3.03} & \underline{23.00} & \underline{2.52} \\
\bottomrule
\end{tabular}}
\end{table}

\begin{table}[t]
\centering
\caption{Effect of training epochs (pref pair: $Syn^{+}$, $Syn^{-}$).}
\label{tab:epoch}
\begin{tabular}{ccccc}
\toprule
\textbf{Loss} & \textbf{Epoch} & \textbf{NV-CER} & \textbf{PCER} & \textbf{CER} \\
\midrule
\multirow{2}{*}{DPO}
& 1 & \textbf{2.59} & \textbf{21.33} & \textbf{2.09} \\
& 2 & 31.32 & 56.89 & 30.83 \\
\midrule
\multirow{3}{*}{DPO+SFT}
& 1 & \underline{3.03} & 23.00 & \underline{2.52} \\
& 2 & 3.25 & \underline{21.78} & 2.78 \\
& 3 & 3.29 & 21.89 & 2.83 \\
\bottomrule
\end{tabular}
\end{table}

\begin{table}[t]
\centering
\caption{Effect of $w_{\mathrm{NV}}$ on NV-CER preference signals.}
\label{tab:weight}
\resizebox{\linewidth}{!}{
\begin{tabular}{ccccccc}
\toprule
\textbf{$w_{\mathrm{NV}}$} &
\textbf{NV-CER} &
\textbf{PCER} &
\textbf{CER} &
\textbf{DNSMOS} &
\textbf{Sim} \\
\midrule
1  & \textbf{2.59}    & 21.33             & \textbf{2.09}    & \textbf{3.346}/\textbf{3.607}/\textbf{4.091} & \textbf{0.891} \\
5  & 3.02             & \underline{18.22} & 2.60             & 3.337/3.601/4.086                            & 0.890 \\
10 & \underline{2.91} & \textbf{18.11}    & \underline{2.52} & \underline{3.339}/3.601/\underline{4.089}    & 0.890 \\
\bottomrule
\end{tabular}}
\end{table}

\textbf{Perceptual quality.}
DNSMOS P.835 \cite{reddy2022dnsmos} predicts speech (SIG), background (BAK), and overall (OVRL) quality, while UTMOS predicts overall MOS for synthesized speech.

\textbf{Speaker similarity.}
Speaker similarity is evaluated using speaker embedding cosine similarity (SECS), computed from speaker embeddings extracted with Resemblyzer\footnote{\href{https://github.com/resemble-ai/Resemblyzer}{https://github.com/resemble-ai/Resemblyzer}}.

\subsubsection{LLM-based multi-rater}
We use Gemini-2.5-Pro as an LLM-based evaluator to assess NV accuracy (\textbf{A}), NV perceptual effect (\textbf{P}), overall naturalness (\textbf{N}), overall quality (\textbf{Q}), and overall expression (\textbf{E}), following the NVVSpeech Challenge Track 2 rubrics\cite{xue2026nvvspeech}. Systems for each utterance are jointly presented with anonymized labels and randomized orders to mitigate system-identity and positional biases in comparative judging. Three independent simulated raters evaluate each sample, with scores assigned according to predefined 1--5 rubrics (0--5 for NV-specific criteria when no NV is audible).

\subsubsection{Subjective Evaluation}
We further conduct an A/B/Tie preference test between the SFT baseline and preference model on the multi-label evaluation set of NV-Bench, providing a human reference for comparison with the objective and LLM-based evaluations. Eleven native-speaking volunteers evaluate four criteria: NV accuracy (NV Acc.); NV naturalness (NV Nat.), with an N/A option when NVs are not comparable (e.g., when an NV is missing); lexical accuracy (Lex. Acc.), assessing text correctness excluding NVs; and overall naturalness (Overall Nat.). The workload is distributed such that each sample is independently evaluated by five volunteers. All participants provided informed consent prior to the evaluation.

\subsection{Implementation Details}

We first perform supervised fine-tuning (SFT) on the pretrained CosyVoice2-0.5B model using the Emilia-NV dataset \cite{liao2026emilia} for 4 epochs with a learning rate of $1\times10^{-5}$, obtaining an Emilia-NV pretrained model (denoted as \textit{Base} in Table \ref{tab:pref_signal_single}, \ref{tab:pref_signal_multi}) that serves as the common initialization for all subsequent post-training experiments. For post-training, we sampled an approximately 100-hour subset of Emilia-NV by prioritizing utterances containing underrepresented NV classes.

Using the \textit{Base} model, we independently generate 8 candidate utterances for each training sample in the post-training subset, from which preference pairs are constructed. Candidate generation follows the default Repetition Aware Sampling (RAS) strategy in CosyVoice2, with $top_p=0.9$, $top_k=30$, $win\_size=10$, and $\tau_r=0.1$ to encourage sampling diversity. During evaluation, $top_p$ and $top_k$ are reduced to 0.8 and 25, respectively, following the default CosyVoice2 configuration.

\begin{table*}[t]
\centering
\caption{Comparison of loss and preference-signal variants on the single-tag test set. Preference pairs use $Syn^{+}$/$Syn^{-}$.}
\label{tab:pref_signal_single}
\scalebox{0.95}{
\begin{tabular}{llcccccc}
\toprule
\textbf{Loss} &
\textbf{Preference Signal} &
\textbf{NV-CER} &
\textbf{PCER} &
\textbf{CER} &
\textbf{UTMOS} &
\textbf{DNSMOS} &
\textbf{Sim} \\
\midrule
\textit{Base} & -- & 3.47 & 32.78 & 2.74 & 2.01 & 3.356/3.616/4.099 & 0.890 \\
SFT  & -- & 3.62 & 30.78 & 2.93 & 2.00 & 3.353/3.613/4.098 & 0.890 \\
\midrule
DPO &
ASR CER &
2.95 &
25.78 &
2.36 &
2.03 &
3.350/3.609/4.095 &
\underline{0.891} \\

DPO &
NV-CER &
\underline{2.59} &
\textbf{21.33} &
\underline{2.09} &
2.01 &
3.346/3.607/4.091 &
\underline{0.891} \\

DPO &
NV-CER + UTMOS &
\textbf{2.58} &
\underline{22.44} &
\underline{2.09} &
\textbf{2.11} &
\textbf{3.377}/\textbf{3.629}/\underline{4.110} &
\underline{0.891} \\
\midrule
DPO+SFT &
ASR CER &
3.35 &
26.00 &
2.78 &
2.01 &
3.353/3.612/4.099 &
0.889 \\

DPO+SFT &
NV-CER &
3.03 &
23.00 &
2.52 &
2.01 &
3.348/3.608/4.097 &
0.890 \\

DPO+SFT &
NV-CER + UTMOS &
3.07 &
24.89 &
2.56 &
\underline{2.04} &
\underline{3.366}/\underline{3.620}/\textbf{4.111} &
0.889 \\
\bottomrule
\end{tabular}
}
\end{table*}

\begin{table*}[t]
\centering
\caption{Comparison of loss and preference-signal variants on the multi-tag test set. Preference pairs use $Syn^{+}$/$Syn^{-}$.}
\label{tab:pref_signal_multi}
\scalebox{0.95}{
\begin{tabular}{llccccccccccc}
\toprule
\textbf{Loss} &
\textbf{Preference Signal} &
\textbf{NV-CER} &
\textbf{PCER} &
\textbf{CER} &
\textbf{UTMOS} &
\textbf{DNSMOS} &
\textbf{Sim} &
\textbf{A} &
\textbf{P} &
\textbf{N} &
\textbf{Q} &
\textbf{E} \\
\midrule
\textit{Base} & -- & 4.98 & 39.28 & 3.69 & 1.92 & \underline{3.355}/\underline{3.617}/\underline{4.099} & 0.889 & 3.28 & 2.34 & 2.43 & 3.28 & 2.28 \\
SFT  & -- & 5.11 & 35.58 & 3.94 & 1.90 & 3.341/3.609/4.086 & 0.890 & 3.35 & 2.31 & 2.37 & 3.23 &2.22 \\
\midrule
DPO &
ASR CER &
4.94 &
37.71 &
3.64 &
\underline{1.95} &
3.342/3.605/4.089 &
\underline{0.891} &
3.45 &
2.73 &
2.85 &
3.59 &
2.67 \\

DPO &
NV-CER &
\underline{4.29} &
\underline{28.40} &
\underline{3.52} &
1.93 &
3.346/3.610/4.091 &
\underline{0.891} &
\underline{3.70} &
2.85 &
2.92 &
3.61 &
2.75 \\

DPO &
NV-CER ($w_{\mathrm{NV}}=10$) &
4.57 &
\textbf{24.58} &
3.91 &
1.94 &
3.352/3.613/4.096 &
0.890 &
\textbf{3.73} &
\textbf{2.91} &
\underline{2.95} &
\underline{3.62} &
\textbf{2.80} \\

DPO &
NV-CER + UTMOS &
\textbf{4.25} &
31.31 &
\textbf{3.27} &
\textbf{2.05} &
\textbf{3.392/3.643/4.119} &
\underline{0.891} &
3.61 &
\underline{2.90} &
\textbf{2.97} &
\textbf{3.72} &
\underline{2.79} \\
\bottomrule
\end{tabular}}
\end{table*}

\begin{table}[htbp]
\centering
\caption{Per-class NV tag retention ratio ($\uparrow$) and false positive count ($\downarrow$) for the SFT and DPO (NV-CER+UTMOS) models on the singe-tag test set (50 utt. per NV type).}
\label{tab:nv_tag_comparison}
\footnotesize 
\renewcommand{\arraystretch}{1} 
\setlength{\tabcolsep}{2.5pt}
\resizebox{\columnwidth}{!}{%
\begin{tabular}{l c c c c | l c c c c}
\toprule
\multirow{2}{*}{\textbf{NV Type}} & \multicolumn{2}{c}{\textbf{Retention}} & \multicolumn{2}{c|}{\textbf{FP}} & \multirow{2}{*}{\textbf{NV Type}} & \multicolumn{2}{c}{\textbf{Retention}} & \multicolumn{2}{c}{\textbf{FP}} \\
\cmidrule(lr){2-3} \cmidrule(lr){4-5} \cmidrule(lr){7-8} \cmidrule(lr){9-10}
& \textbf{SFT} & \textbf{DPO} & \textbf{SFT} & \textbf{DPO} & & \textbf{SFT} & \textbf{DPO} & \textbf{SFT} & \textbf{DPO} \\
\midrule
Breathing & \textbf{0.68} & 0.52 & 63 & \textbf{37} & Question-oh & 0.66 & \textbf{0.68} & 1 & \textbf{0} \\
Confirm-en & 0.94 & \textbf{0.96} & 46 & \textbf{35} & Question-yi & 0.76 & \textbf{0.88} & \textbf{0} & 1 \\
Cough & 0.72 & \textbf{0.86} & 1 & \textbf{0} & Shh & 0.74 & \textbf{0.80} & 1 & \textbf{0} \\
Crying & 0.00 & 0.00 & 0 & 0 & Sigh & 0.78 & \textbf{0.94} & 6 & \textbf{4} \\
Dissat-hnn & 0.86 & \textbf{0.94} & \textbf{2} & 5 & Surprise-ah & 0.64 & \textbf{0.82} & 10 & \textbf{5} \\
Laughter & 0.92 & \textbf{0.96} & 10 & \textbf{3} & Surprise-oh & 0.88 & \textbf{0.98} & 22 & \textbf{19} \\
Question-ah & \textbf{0.88} & 0.86 & 8 & \textbf{0} & Surprise-wa & 0.90 & \textbf{0.96} & \textbf{1} & 2 \\
Question-ei & 0.98 & 0.98 & 7 & \textbf{2} & Surprise-yo & 0.88 & \textbf{0.98} & 0 & 0 \\
Question-en & 0.80 & \textbf{0.88} & 11 & \textbf{1} & Uhm & \textbf{0.76} & 0.66 & 22 & \textbf{10} \\
\bottomrule
\end{tabular}
}
\end{table}

\begin{table}[t]
\centering
\caption{Human Preference on the multi-tag test set.}
\label{tab:human_preference}
\resizebox{\columnwidth}{!}{%
\begin{tabular}{lcccc}
\toprule
\textbf{Criterion} &
\textbf{Preference} &
\textbf{Tie} &
\textbf{SFT} &
\textbf{N/A} \\
\midrule
NV Acc.
& 340 (17.39\%)
& 1342 (68.64\%)
& 273 (13.96\%)
& -- \\

NV Nat.
& 594 (30.38\%)
& 669 (34.22\%)
& 584 (29.87\%)
& 108 (5.52\%) \\

Lexical Acc.
& 186 (9.51\%)
& 1630 (83.38\%)
& 139 (7.11\%)
& -- \\

Overall Nat.
& 674 (34.48\%)
& 598 (30.59\%)
& 683 (34.94\%)
& -- \\
\bottomrule
\end{tabular}%
}
\end{table}

Unless otherwise specified, all post-training experiments use an LR of $1\times10^{-6}$, dynamic batching with a maximum of 1000 frames per batch, gradient accumulation of 40, one training epoch to ensure consistent comparisons and minimize training instability, and $\beta=0.75$. We adopt the NV-ASR model from \cite{ni2026nv} to compute NV-CER, which fine-tunes SenseVoice-Small \cite{an2024funaudiollm} on multiple public NV speech corpora \cite{liao2026emilia, borisov2025nonverbaltts, wang2024disfluencyspeech, ye2025scalable, wu2025smiip, mai2026mnv}. It recognizes all 18 NV tags in Emilia-NV, with the majority (two-third) achieving around 90\% or higher retention (i.e., the proportion of reference tags correctly identified).

\subsection{Design Choices}
\label{subsubsec: design choices}

\textbf{Loss formulation:}
We investigate whether combining DPO with the supervised fine-tuning (SFT) objective improves performance and training stability. The SFT loss is weighted by 0.2 to keep its magnitude comparable to the DPO loss.

\textbf{Preference pair construction:}
We investigate how preference-pair construction affects optimization. For preferred responses, we compare three strategies: (1) the default, selecting the top-ranked synthetic candidate (Section~\ref{subsubsec: NV-Aware Preference Pair Construction}); (2) always selecting ground-truth speech; and (3) selecting ground-truth speech when its NV-CER is lower than the top-ranked synthetic candidate, otherwise using the synthetic candidate. For rejected responses, we additionally consider synthesized speech without NVs, following \cite{bai2026synparaspeech}.

\textbf{Preference signal:}
Unless otherwise stated, preference scores are computed using NV-CER (Eq. \ref{eq:nv_cer}) with $w_{\mathrm{NV}}=1$. We investigate three preference-signal alternatives: (1) conventional pinyin CER from the general-purpose SenseVoice-Small ASR model \cite{an2024funaudiollm}; (2) varying $w_{\mathrm{NV}}$ in Eq \ref{eq:nv_cer_w} to explicitly control the relative importance of NV errors; (3) a composite NV recognition and perceptual quality score, defined as:
\[
s=\mathrm{CER}_{\mathrm{NV}}-\lambda(\mathrm{UTMOS}-1)/4,\quad \lambda=0.4.
\]

\section{Results and discussion}
\label{sec:result}

\subsection{Loss and Preference Pair Ablation}

We investigate loss formulation and preference-pair construction on the single tag testset of NV-Bench. As shown in Table~\ref{tab:ablation_setting}, the $Syn^{+}$/$Syn^{-}$ pair achieves the best performance and is adopted in subsequent experiments. Using GT/Hybrid in the preferred response reduces performance, while $Syn^{\mathrm{NoNV}}$ also underperforms $Syn^{-}$ because the base model rarely omits NVs, making NV-free speech weak negative examples.

Table~\ref{tab:ablation_setting} and ~\ref{tab:epoch} show that DPO alone is unstable, with performance deteriorating after the first or second epoch. This instability stems from policy drift away from the reference distribution, causing substantial sequence-level likelihood degradation, with rejected speech sequences experiencing a larger decline. Adding SFT substantially improves training stability but weakens the DPO effect for $Syn^{+}$/$Syn^{-}$. We therefore use DPO in subsequent experiments and include DPO+SFT in selected experiments as a robustness check.

\subsection{Investigation on the impact of NV-Weight tuning}
Table~\ref{tab:weight} evaluates the effect of the NV tag weight on the single tag testset. Increasing $w_{\mathrm{NV}}$ improves PCER, with a modest increase in CER. Across all settings, CER remains below that of the Base and SFT models in Table~\ref{tab:pref_signal_single}. This result suggests that weighted NV-CER can modulate the relative emphasis on NV-related and lexical errors without modifying the underlying preference optimization framework.

\subsection{Comparison with Baselines}

Table~\ref{tab:pref_signal_single} compares NV-aware preference signals with conventional ASR-CER preference signal, the \textit{Base} model, and a SFT model (continued SFT training on \textit{Base} with the post-training subset, identical LR and training steps as the DPO experiments) on the single-tag test set. SFT slightly degrades performance from \textit{Base}, whereas NV-CER consistently improves both lexical and NV-related metrics and outperforms conventional ASR-CER. Combining NV-CER with UTMOS achieves similar transcription accuracy while slightly improves UTMOS and DNSMOS. The same trends hold for both DPO and DPO+SFT, demonstrating robustness across loss formulations. Speaker similarity remains largely unchanged.

Table \ref{tab:pref_signal_multi} further evaluates the methods on the multi-tag test set using objective and LLM-based metrics. NV-CER-based preference signals consistently outperform the baselines.

Table \ref{tab:nv_tag_comparison} compares DPO against SFT across individual NV categories on the single-tag test set. DPO improves most NVs, increasing retention ratios while reducing or maintaining FP. Categories with lower retention under DPO consistently show significant FP reduction. \textit{Breathing} exhibits the highest FP due to unannotated acoustic breathing in the training transcripts, while \textit{Crying} is not learned by either model due to data sparsity (only 1,288 utterances in Emilia-NV).

\subsection{Human Evaluation}

Table \ref{tab:human_preference} compares the preference model (NV-CER+UTMOS, Table \ref{tab:pref_signal_multi}) with the SFT baseline via subjective listening tests. High tie rates for NV Acc. and Lex. Acc. are consistent with the strong baseline performance reflected by the CER-based metrics. Among non-tied judgments, the preference model achieves statistically significant gains over the SFT baseline for NV Acc. (55.5\% vs. 44.5\%; $p = 0.0076$) and Lex. Acc. (57.2\% vs. 42.8\%; $p = 0.0106$), with $p$-values derived from two-sided binomial tests, aligning with objective and LLM evaluations. No obvious preference is observed for NV Nat. or Overall Nat., suggesting that the accuracy gains do not compromise perceived naturalness. This aligns with the preference signal, which targets NV and lexical correctness through NV-CER, while UTMOS does not explicitly assess NV naturalness. The discrepancy with LLM evaluation further highlights the need for better metrics to distinguish NV naturalness when perceived differences are small. Overall, NV-aware preference signals improve targeted accuracy while preserving perceived naturalness.

\vfill\pagebreak
\clearpage

\footnotesize

\bibliography{refs}

\end{document}